\documentstyle[aps,twocolumn,epsf]{revtex}
\begin{document}
\draft
 \title{
Phase diagram of the metal-insulator transition in 2D electronic systems}
\author{J.S. Thakur and D. Neilson}
\address{
School of Physics, The University of New South Wales, Sydney 2052, Australia.
\\[3pt]
\  \\ \medskip}\author{\small\parbox{14cm}{\small
We investigated the interdependence of the effects of disorder and
carrier correlations on the metal-insulator transition in
two-dimensional electronic systems. We present a quantitative
metal-insulator phase diagram.  Depending on the carrier density we
find two different types of metal-insulator transition -- a continuous
localization for $r_{s}\lesssim 8$ and a discontinuous transition at
higher $r_s$. The critical level of disorder at the transition
decreases with decreasing carrier density. At very low carrier
densities we find that the system is always insulating. The value of
the conductivity at the transition is consistent with recent
experimental measurements. The self-consistent method which we have
developed includes the effects of both disorder and correlations on the
transition, using a density relaxation theory with the Coulomb
correlations determined from numerical simulation data.
 \\[3pt]{PACS numbers:
 73.20.Dx,71.30.+h,73.40.-c}
}}\address{} \maketitle

\narrowtext

A metal-insulator transition (MIT) has now been observed in a number of
different 2D electronic systems over a wide range of carrier densities
and levels of disorder. At the transition the carrier density parameter
$r_{s}$, which measures the strength of the carrier correlations,
covers values $7 \alt r_{s}\alt35$. \cite{Kravchenko,SHG,MDH} The
strength of disorder at the transition also covers a wide range, and
depends on the value of the critical $r_{s}$. In Si/SiGe\ \ \cite{MDH}
the transition occurs at low carrier densities and high mobilities.
For the transition in Si MOSFETs which occurs at higher densities, the
mobilities are much smaller.  A recent experimental phase diagram
\cite{Hamilton} provides a relation between the strength of the
correlations and the disorder.  It shows that the critical level of the
disorder at the transition diminishes as the correlations grow
stronger.

In spite of a great deal of experimental and theoretical work the
nature of the conducting state remains unclear and controversial. Unlike
conventional metals, its Coulomb interaction energy is typically an
order of magnitude larger than its Fermi energy. Numerous proposals
have been made about the causes of the stabilization of this conducting
phase. These include strong Coulomb repulsions between the carriers
\cite{TN1,pikus}, or an anomalous enhancement of the spin-orbit
interaction due to the broken inversion symmetry of the confining
potential well in Si MOSFETs \cite{pudalov}.  References \cite{PWI}
proposed that the conducting phase is superconducting, with the pairing
of the carriers being mediated by the dynamic correlation hole
surrounding each carrier.  Thakur and Neilson \cite{TN2} demonstrated
the existence of superconducting pairing due to strong Coulomb
correlations in the presence of disorder.  They found up to levels of
disorder typical of high quality Si MOSFETs that the superconductivity
persists.

The conducting phase is destroyed not only at low but also at high
carrier densities.  Kravchenko {\it et al}\ \cite{Kravchenko}
originally observed the insulator-to-metal transition by increasing the
carrier density. Recently it has been reported that if the carrier
density is increased further there is an upper critical value where the
conducting phase again becomes unstable and the system enters another
insulating phase \cite{Hamilton}. This insulating phase has the
characteristics of a single-particle localized state, while the
insulating phase at low densities is a coherent insulator with
properties similar to a Wigner crystal or glass.  The critical density
for the second transition, typically around $r_{s}\alt7$, is still not
low enough for electron interactions to be neglected. In this paper we
treat Coulomb interactions and disorder on an equal footing in order to
investigate the transitions at both small and large $r_s$.

In the strong correlation limit and in the absence of disorder,
electrons localize to form a Wigner solid with long range crystalline
order.  Here we consider interacting charge carriers in the presence of
weak disorder which would destroy any long range order.  We have
previously proposed that strong Coulomb correlations in the presence of
disorder can localize electrons into a coherent glassy insulator with
liquid-like short-range order \cite{TN1}.  We studied the
metal-insulator transition in terms of the ergodic to nonergodic
transition of the charge density fluctuations.  The possibility of
localization into such a glassy state has also been discussed recently
by Chakravarty {\it et al} \ \cite{Chakravarty}.

We define the order parameter for the glass state as
$f(q)=\lim_{t\rightarrow \infty }\Phi (q,t)$, where $\Phi (q,t)\equiv
\bbox (N(q,t)\bbox \mid N(q,0)\bbox )$ is the Kubo-relaxation function
for the dynamical density variable $N(q,t)$.  The normalized variable
$N(q,t)={{\rho (q,t)}/{\sqrt{\chi (q)}}}$, where $ \rho (q,t)$ is the
usual density fluctuation operator and $\chi (q)$ is the static
susceptibility. When the order parameter is non-zero, spontaneous
density fluctuations do not decay at infinite time 
and the system will be an insulator.  Conversely if $f(q)$ is zero then
our system is in a conducting phase.  Since our order parameter would
be zero for any conducting phase it gives no indication about the
precise nature of the conducting phase.

Within the Mori-Zwanzig formalism \cite{Mori} $\Phi (q,t)$ is
calculated in terms of the memory function $M(q,t)$, which we evaluate
using mode-coupling theory \cite{MC}. In the limit $t\rightarrow \infty
$ the relaxation function reduces to
\begin{equation}
f(q)=\frac{1}{1+{\Omega (q)}/{M(q)}}\ ,  
\label{z->0f(q)}
\end{equation}
where $\Omega (q)={q^{2}}/(m^{\star }\chi (q))$ and $M(q)=\lim_{t\rightarrow
\infty }M(q,t)$. $m^\star$ is the carrier effective mass.  We express 
$M(q)=M_{cc}(q)+M_{ic}(q)$, where 
\begin{eqnarray}
M_{cc}(q) &=&\frac{1}{2m^{*}q^{2}}\sum_{q^{\prime }}\left[ V(q^{\prime })( 
\bbox q\cdot \bbox q^{\prime })+V(|\bbox q-\bbox q^{\prime }|)\right.
\nonumber \\
    &\times&\left.(\bbox q\cdot
( \bbox q-\bbox q^{\prime }))\right] ^{2}\chi (q^{\prime })\chi (|\bbox q-
\bbox q^{\prime }|)f(q^{\prime })f(|\bbox q-\bbox q^{\prime }|)  \nonumber
\\
\nonumber \\
M_{ic}(q) &=&\frac{1}{m^{*}q^{2}}\sum_{q^{\prime }}\left[ n_{i}\langle |U_{
\text{imp}}(q)|^{2}\rangle +\langle |W_{\text{surf}}(q)|^{2}\rangle \right]
\nonumber \\
    &\times&( \bbox q\cdot \bbox q^{\prime })^{2}
\chi (|\bbox q-\bbox q^{\prime }|) f(|\bbox q-\bbox q^{\prime }|).  
\label{M(q)}
\end{eqnarray}
The $M_{cc}(q)$ part of the memory function originates from the
interactions between the carriers, and the $M_{ic}(q)$ from the
carrier-disorder interactions.  $V(q)=2\pi e^{2}/\epsilon q$ is the
Coulomb potential with dielectric constant $\epsilon $.
$U_{\text{imp}}(q)$ is the carrier-impurity potential for randomly
distributed monovalent Coulombic impurities of density $n_i$ which are
embedded in the carrier layer. $W_{\text{surf}}(q)$ is the surface
roughness scattering term. Details of the disorder potentials used are
given in Ref.\ \ \cite{TN1}.

To evaluate the static susceptibility $\chi (q)$ with correlations, we use
the generalized Random Phase Approximation expression,
\begin{equation}
\chi (q)=\chi_{0}(q)/\{1+V(q)(1-G(q))\chi _{0}(q)\},
\end{equation}
where $\chi_{0}(q)$ is the Lindhard function for non-interacting
electrons.  The local field factor $G(q)$ accounts for the correlations
between the carriers.  We evaluate $G(q)$ from ground state properties
of the electron liquid \cite{Ceperley} using the
fluctuation-dissipation theorem \cite{SNS}.

The memory functions $M_{cc}(q)$ and $M_{ic}(q)$ mutually influence
each other through $f(q)$.   The nature of the localized state at the
transition is largely determined by which of these memory functions is
dominant. At low densities and small levels of disorder the
interactions between the carriers dominate and $M_{cc}(q) $ is much
larger than $M_{ic}(q)$. In this case the localization is primarily
caused by many-body effects and the localized state is a coherent
frozen insulator. At the transition the order parameter $f(q)$ jumps
from zero to non-zero and the system undergoes a discontinuous
transition. In contrast, at high densities and high levels of disorder,
where the carrier-disorder scattering dominates, $M_{ic}(q)$ is much
larger than $M_{cc}(q)$. The localization transition in this case is to
a non-coherent state where the carriers localize independently. We find
that this transition is continuous, with the order parameter $f(q)$
continuously increasing with disorder.   Our model thus predicts the
two distinct types of metal-insulator transition.

 \begin{figure}
 \epsfxsize=8.0cm
 \epsffile{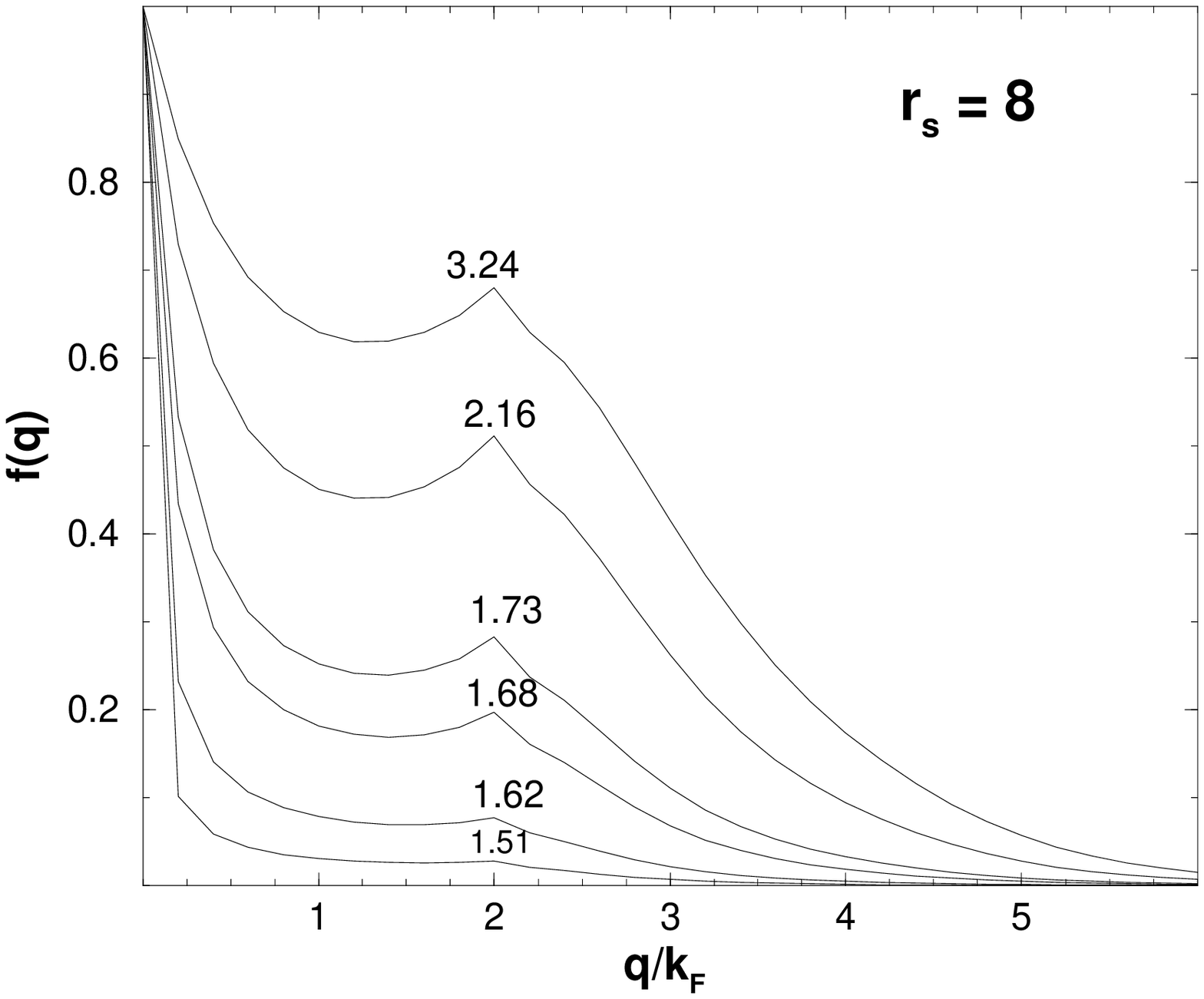}
 \epsfxsize=8.0cm
 \epsffile{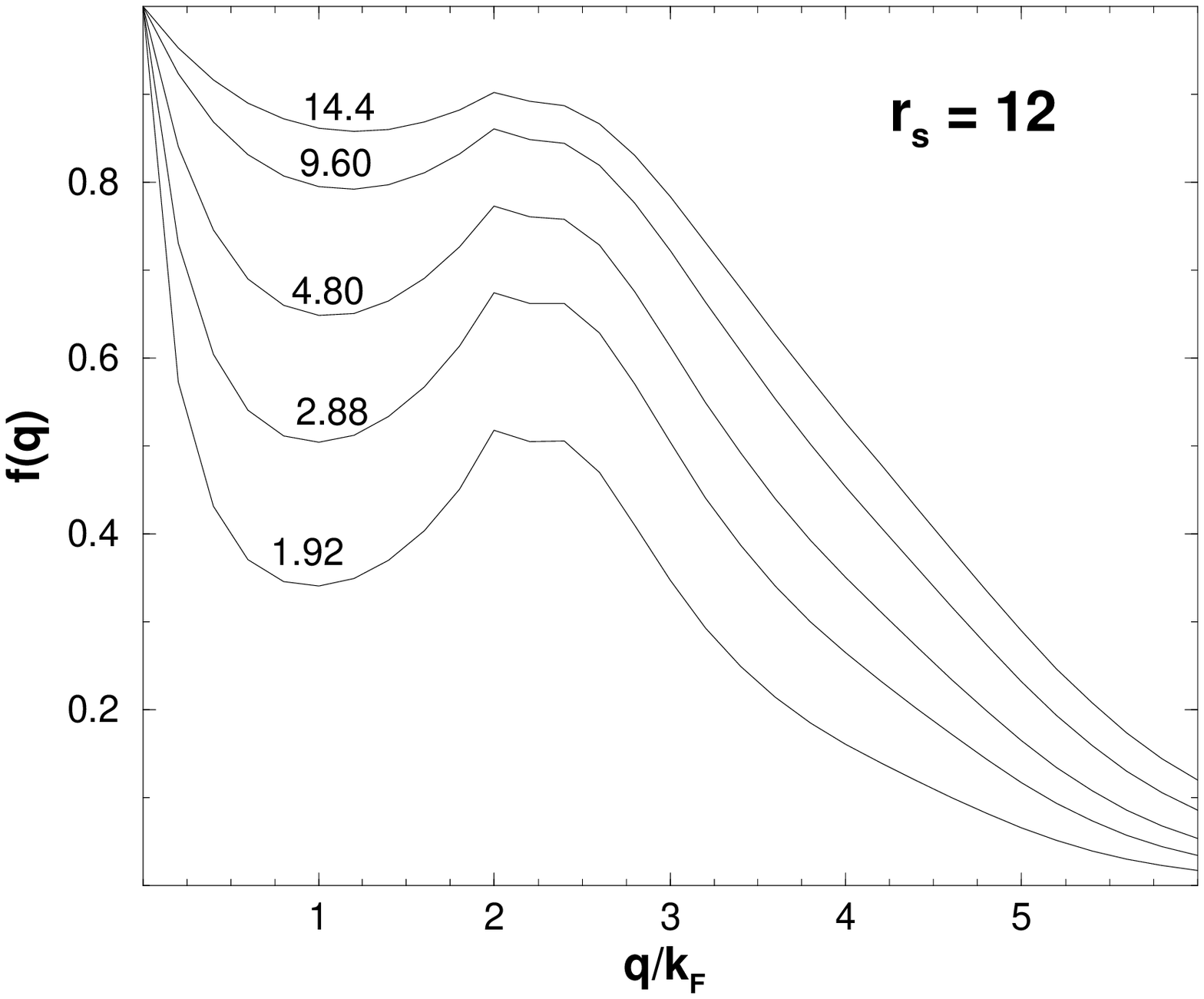}
 \epsfxsize=8.0cm
 \epsffile{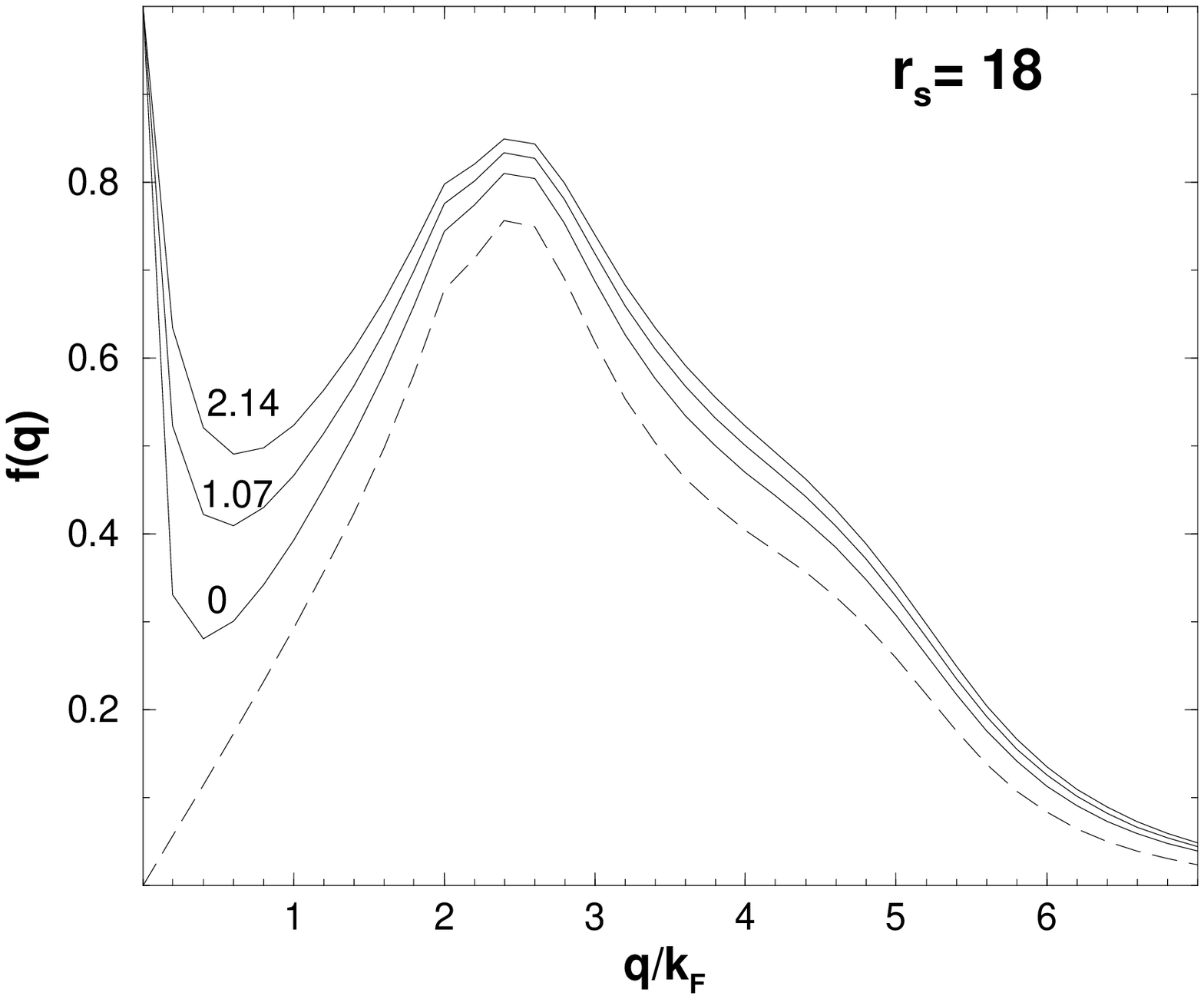}
  \caption[dummy3]{
Order parameter $f(q)$.  Curve labels are impurity densities $n_{i}$.
Surface roughness is included in all cases.\\
 a. $r_{s}=8$.  $n_{i}$ is in units of $10^{10}$cm$^{-2}$. \\
 b. $r_{s}=12$.  $n_{i}$ is in units of $10^{9}$cm$^{-2}$.
$f(q)$ is zero for $n_{i}<1.92\times 10^{9}$cm$^{-2}$.\\
 c. $r_{s}=18$.  $n_{i}$ is in units of $10^{8}$cm$^{-2}$.
Dashed line is $n_{i}=0$ with no surface roughness.
 \label{f(q)}}
\end{figure}

We solve Eqs.\ \ref{z->0f(q)} and \ref{M(q)} self-consistently for the order
parameter $f(q)$. We vary the strength of the disorder potential by varying
the impurity density $n_i$ while keeping the surface roughness at each $r_s$
fixed.

In Fig.\ \ref{f(q)} we show the order parameter $f(q)$ for three values
of $ r_{s}=8$, $12$ and $18$. The multiple curves for each fixed
$r_{s}$ correspond to increasing levels of disorder. At the highest
carrier density, Fig.\ \ref{f(q)}a, the localization is in general
dominated by the $M_{ic}(q) $ part of the memory function. Here $f(q)$
increases continuously with impurity density, indicating continuous
localization.  These characteristic features that $f(q)$ evolves
continuously with the disorder level and is non-zero for any disorder
indicates a localized state where the particles localize independently.
This is similar to Anderson-type localization.  Since $f(q)$ is always
non-zero there is no conducting phase when $r_s=8$.  The peak in $f(q)$
at $q=2k_{F}$ is a result of the well--known cusp in the
two-dimensional $\chi _{0}(q)$ and is of no importance here.

In contrast, in Fig.\ \ref{f(q)}b for $r_{s}=12$ the order parameter is
zero for small non-zero levels of disorder.  This indicates the
existence of a conducting phase. If we increase the level of disorder, a
critical value is passed at which $f(q)$ jumps discontinuously to
non-zero values and at that point there is a metal-insulator transition
\cite{TN1}. The short-range coherent order of the insulator state is
reflected by a peak in $f(q)$ at the reciprocal nearest neighbor
distance, $q\approx 2.4k_{F}$.  As we increase the disorder further,
the $f(q)$ increases continuously.  For very high levels of disorder,
$n_{i}> 6\times 10^{9}$ cm$^{-2\text{ }}$, the overall shape of the
$f(q)$ evolves towards a Gaussian indicating the development of a
non-coherent insulator. We find in the range $ 8\lesssim r_{s}\lesssim
18$ that the critical level of disorder required to localize the
carriers decreases with increasing $r_{s}$ and that the non-coherent
state also occurs at decreasing $n_{i}$.

By $r_{s}=18$  the system localizes without disorder and there is no
conducting phase. This large $r_{s}$ localization is driven purely by the
correlations between the carriers. In Fig.\ \ref{f(q)}c $f(q)$ is
non-zero for  $n_{i}=0$ and no surface roughness scattering (dashed
line). Without disorder the $f(q)$ goes to zero as $q$ goes to zero.
The solid lines show $f(q)$ with surface roughness scattering
included.

\begin{figure}
 \epsfxsize=9.5cm
 \epsffile{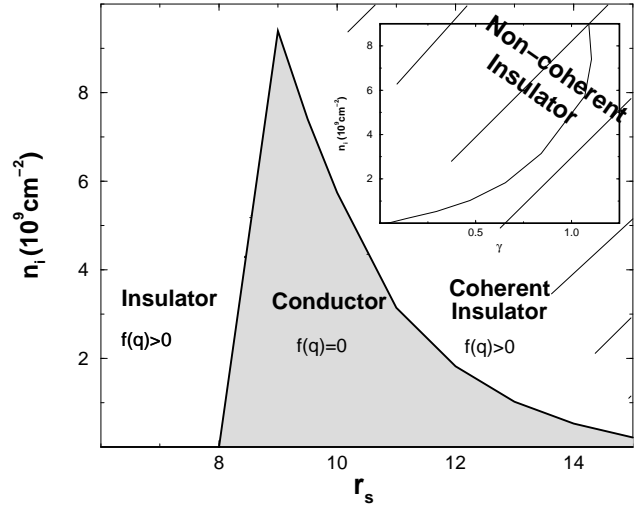}
  \caption[dummy3]{
Zero temperature phase diagram. Axes are  impurity density $n_{i}$ and
carrier density parameter $r_{s}$. In the insulating phases the order
parameter $f(q)>0$. In the conducting phase $f(q)=0$. The transition at
higher densities is continuous. When $r_{s}>9$ the $f(q)$
discontinuously jumps from zero to non-zero values at the transition.
Inset shows the corresponding scattering rate $\gamma/(2E_F)$ at the
transition as a function of $n_{i}$.
 \label{phasespace}}
 \end{figure}

The property that the critical level of disorder for the
metal-insulator transition is dependent on the carrier density  is
associated with the changing strength of the correlations.  The zero
temperature phase diagram in Fig.\ \ref{phasespace} shows the
relationship between the carrier density and the critical level of
disorder at the transition. Disorder includes both the impurity density
$n_{i}$ and surface roughness. For $r_{s}\alt8$  the system is
insulating in the presence of any disorder, and the order parameter
$f(q)$ is always non-zero. For $8< r_{s}<18$ the order parameter is
zero for $n_{i}$ less than a critical value and we have the conducting
phase. At the right phase boundary the transition to the insulating
phase is discontinuous and $f(q)$ jumps discontinuously to non-zero
values as we cross it. The critical level of disorder for the
transition decreases with decreasing density. Near the phase boundary
where the disorder level is slightly larger than the critical value the
system is in the coherent insulating state.  For very large disorder
the $f(q)$ has evolved into a Gaussian-like shape (see
Fig.\ \ref{f(q)}b) and the system becomes a non-coherent insulator.
For $r_{s}\gtrsim 18$ the order parameter is non-zero without disorder
and the conducting phase has disappeared. Our phase diagram has common
features with the conceptual phase diagram discussed by Chakravarty
{\it et al} \ \cite{Chakravarty}.

The inset in Fig.\ \ref{phasespace} shows the relation between $n_{i}$ and
the scattering rate $\gamma $ off the disorder. Within the memory function
formalism the non-linear equation for $\gamma $ is given by \cite{Gotze} 
\begin{eqnarray}
{\rm i}\gamma &=&-\frac{1}{2m^{\star }n_{c}}\sum_{q}q^{2}\left[ n_{i}\langle
|U_{\text{imp}}(q)|^{2}\rangle +\langle |W_{\text{surf}}(q)|^{2}\rangle
\right] \nonumber \\
   &\times&\left( \frac{\chi (q)}{\chi ^{(0)}(q)}\right) ^{2}\frac{\phi _{0}(q,
{\rm i}\gamma )}{1+{\rm i}\gamma \phi _{0}(q,{\rm i}\gamma )/\chi ^{(0)}(q)}
\ ,
\end{eqnarray}
where $\phi _{0}(q,{\rm i}\gamma )=(1/{\rm i}\gamma )\left[ \chi ^{(0)}(q,
{\rm i}\gamma )-\chi ^{(0)}(q)\right] $ is the relaxation spectrum for
non-interacting carriers that scatter off the disorder.
Thakur and Neilson \cite{TN2} showed for the density range $5\leq
r_s\leq10$ that the superconducting phase persists at least up to
values of $\gamma/(2E_F)=1$.  They also showed that the effective
potential becomes more strongly attractive as $r_s$ increases.  This
suggests that our conducting phase in Fig.\ \ref{phasespace}, which is
only found for $\gamma/(2E_F)\alt1$, is a superconductor.

\begin{figure}
 \epsfxsize=8.cm
 \epsffile{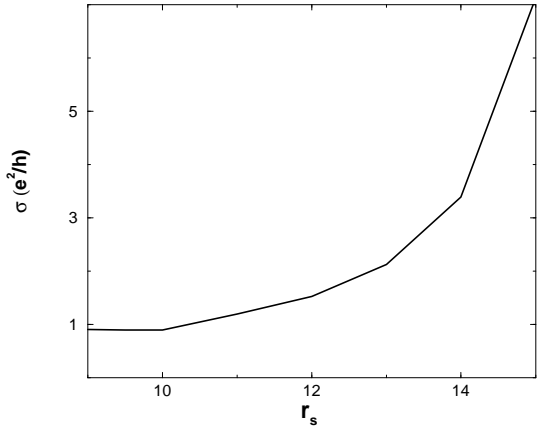}
  \caption[dummy3]{
Conductivity $\sigma $ at the critical disorder for the transition
as a function of $r_{s}$.
 \label{sigma}}
\end{figure}

Using the Drude model, $\gamma $ can be related to the zero temperature
conductivity, $\sigma =(ne^{2}/m^{\star })\gamma ^{-1}$. In
Fig.\ \ref{sigma} we show $\sigma $ at the transition as a function of
$r_{s}$. In the range $ 9< r_{s}<14$, $\sigma $ lies between $1\alt
\sigma/(e^{2}/h) \alt3$. This is consistent with experimental values.
At lower densities the $\sigma $ calculated within the Drude model
increases rapidly.  This is associated with the decrease in the
critical level of disorder at the transition.  Since the Drude
approximation is inapplicable for large $r_s$ this increase in $\sigma$
at low density should not necessarily be seen experimentally.

\acknowledgements

This work is supported by Australian Research Council Grant.  We thank
Philip Philips for useful comments.

\end{document}